# Excimer laser cleaning of black sulphur encrustation from silver surface

Mohammad Shahid Raza, Sankha Shuvra Das, Parimal Tudu, Partha Saha*
Department of Mechanical Engineering, Indian Institute of Technology Kharagpur, West Bengal, India-721302
*Email id: p.saha@iitkgp.ac.in

**Abstract**

The process of localized cleaning of silver sulphide from the silver surface has been investigated in this work. An artificial black encrustation was generated on the silver surface and its laser cleaning was performed using an excimer laser (KrF based) working at a wavelength of 248 nm and pulse width 25 ns. Laser process parameters i.e. laser fluence and number of pulses were used to perform the experiment and after different optical, microstructural, elemental, microhardness and surface topography analysis of the laser ablated zone was performed. The parametric window selected for the excimer laser cleaning were laser fluence (200-400 mJ/cm$^2$), repetition rate 4 Hz and number of pulses (60-300). It was observed that excimer laser is very effective in the cleaning of black encrustation having ~ 6 wt. % of sulphur from the surface without any significant removal of the silver substrate. The fluence value of 260.48 mJ/cm$^2$ was observed as above threshold value for removal of sulphide from the encrusted surface while a fluence of 384.85 mJ/cm$^2$ and 220 pulses showed the maximum removal of sulphur encrustation (~0.38 wt. % of sulphur remaining). Formation of the peripheral rim was observed to occur above 280.17 mJ/cm$^2$ of fluence value. The topography of the ablated surface showed a removal of 5-10 µm of the sulphide layer from the surface whereas the thickness of the sulphide coating was around 8-10 µm. Microhardness value showed a change of microhardness of the silver surface from 75.12 HV$_{0.5}$ (uncoated parent silver) to 96.07 HV$_{0.5}$ (centre of the ablated zone).Finally, it was concluded that excimer laser cleaning is a novel method for localized removal of silver sulphide encrustation from silver surface.

# 1. Introduction

Silver is a metal which is gaining wide importance in research due to its increasing application areas. Starting with aesthetic values, it is used in electronics industry, medicines, equipment, coins, water purifiers, containers, etc. Among them, electronics industries are using silver to make printed circuit boards, connectors, speaker wires and superconductors [1]. However, it is seen that silver gets tarnished with time and it forms a black layer on its surface. This causes a change in its morphological, electrical and thermal properties. Silver surface black encrustation is basically silver sulphide formed by reaction of the silver with sulphur compounds present in the atmosphere. With time this tarnish (also called as encrustation) develops a dark colour coating on the silver surface. Removal of this coating is very important in order to maintain the aesthetic as well as monetary value of silver.

While comparing laser cleaning with other approaches like mechanical and chemical procedures[2], it was observed that the controlled removal of the substrate through the mechanical process is very difficult. Silver sulphide formed on the surface does not get dissolved with most of the alkalis and acids. Also, the contaminants used in chemical cleaning possess difficulties for application zones like in circuit boards and electronic components. As lasers are a highly clean source of energy, so, laser cleaning is the most suitable method for such type applications.

Research on laser cleaning dates back in the seventies when it was first used to clean the stone arts of the human being. The preliminary studies on various laser cleaning techniques that can be applied for surface contaminants showed that pulse laser cleaning is an effective tool for removing the contaminants while preserving the substrate from getting removed [3].The major research fields of laser cleaning were basically concentrated on cleaning of ancient scriptures and monuments, the contaminated surface of materials and automobile components. Different laser classes were used to clean the encrustation from the ancient pictures and dogeourotypes [4–8].Excimer pulsed laser was used for the cleaning of ancient manuscripts made up of biological composite materials [9].The study provided some insight of the fundaments of laser cleaning. Different excimer lasers like KrF based and XeCl based lasers were also compared for removal of encrustation using characterisation techniques like Fourier transform infrared spectroscopy

(FTIR), scanning electron microscopy (SEM), Energy dispersive X-ray analysis (EDX), X-ray diffraction (XRD), laser-induced breakdown spectroscopy (LIBS) and optical microscopy (OM) [10].These characterisation techniques were also applied to study the colour changes in marbles during Nd: YAG laser cleaning. They found that the second or third harmonics were showing a better result with respect to first harmonics [11].This research broadened the application areas of Nd: YAG laser in the cleaning of contaminants which was later applied for cleaning of soiled cellulose paper using 532 nm Nd: YAG laser [12].

Laser cleaning was also applied on copper to clean the electronic circuit boards using excimer laser [13] and the effect of process parameters on cleaning was also studied with Nd: YAG lasers [14] and femtosecond pulsed lasers [15].Q- switched Nd: YAG lasers were used for oxide removal from copper surfaces [16] where the effect of different wavelengths on the material removal amount and the material removal mechanism was studied. Since the effect of high ablation rate due to liquid film on the surface was already known, so researchers also tried to understand the mechanism of steam laser cleaning and its effectiveness which showed that efficiency of cleaning can be increased using thin liquid film on the surface [17]. Some new mechanisms related to Nd: YAG laser cleaning was also tried [18]. Cleaning of silver using Nd: YAG laser was performed for a wide range of wavelengths which showed that the smallest wavelength used (266 nm) was most effective to clean silver without any damage [19]. Gold films having carbon layers were cleaned using pulsed Nd: YAG laser [20].

Feasibility study of the use of high power diode laser for paint removal was done in 2003 where the removal of chlorinated rubber (CR) and epoxy resin paints from surfaces were performed [21]. Similarly, the effect of high repetition rate as well as the laser operating mode on ablation of paint was also reported respectively [22,23]. The $CO_2$ laser was also used for laser cleaning operation on titanium alloys [24]. Surface cleaning of diesel engine parts was also reported where the laser was used successfully to clean the carbon deposits on the plug region and engine valves respectively [25,26]. Apart from experimental approaches, modelling and simulations were also performed to understand the physics and mechanism of laser cleaning [27,28].

Previous studies have shown that excimer laser has got a vast potential in localised cleaning operations. Laser ablation mechanisms and their conditions to take place depending on pulse time, Fluence and Intensity of the pulses [29]. Various thermal properties that differentiate silver sulphide from the silver surface were already studied and well documented [7] which shows the need to remove the sulphur from the silver surface for the proper functioning of the silver components. But the research on localised excimer laser cleaning in removing sulphur encrustation from a silver surface has not been characterized and analysed. Some of the critical issues like threshold energy value for cleaning, Cleaning mechanism and formation of the peripheral rim (a common phenomenon in ablation) needs proper attention. The aim of the present research is to study the effect of excimer laser input parameters (such as laser fluence and number of pulses) on wt. % removal of silver sulphide from the silver surface using optical, topographical, microstructural and microhardness analysis. This study is important because the controlled removal of sulphur without removing the silver requires proper use of process parameter window.

## 2. Materials and Methods

The silver sheet of dimensions 7.5 mm × 7.5 mm × 2.2 mm was used for the experimentation. Table 1 gives the EDS analysis of the silver surface and the encrusted sample surface. Silver (~97.18 wt. %) with about 0.07 wt. % sulphur which after encrustation gets increased up to 5.56 wt. %. was used for the experimentation.

**Table 1** Elemental composition of non-tarnished silver sample and encrusted sample

| Element | C | O | S | Cu | Ag |
|---|---|---|---|---|---|
| Wt. %. Non tarnished silver | 1.91 | 0.12 | 0.07 | 0.72 | 97.18 |
| Wt. %. tarnished silver | 4.04 | 3.86 | 5.56 | 1.68 | 84.87 |

*2.1 Generation of sulphur encrustation*

In order to produce the silver sulphide encrustation, Kipp's apparatus has been used. Kipp's apparatus setup (as shown in fig. 1) is a reliable media to transfer various gases on the surface. Generally, silver gets tarnish if it exposes to open environment (by reacting with hydrogen sulfide). The standard lab preparation of hydrogen sulfide gas can be a reaction of ferrous sulfide with hydrogen chloride (HCl) inside Kipp's apparatus. Further, the generated hydrogen sulphide was made to react with silver to form silver sulphide (eq. 1).

$$FeS(s) + 2\,HCl(l) = FeCl_2 + H_2S(g) \qquad (1a.)$$

$$2Ag(s) + H_2S(g) = Ag_2S(s) + H_2(g) \qquad (1b.)$$

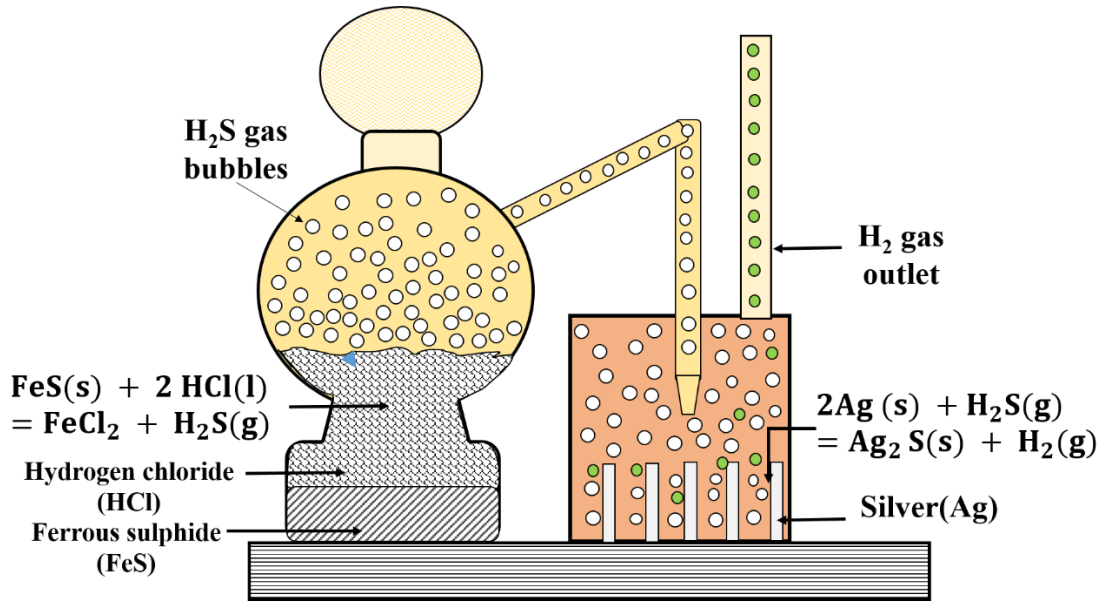

**Fig. 1.** Schematic of Kipp's apparatus arrangement used to create silver sulphide encrustation.

After the generation of silver sulphide on the surface, trial experiments were carried out. Optical images of the encrusted samples were taken to observe the tarnish developed on the surface. It was observed that a dull layer was formed with combined dark patches and light grey colour areas. The amount of encrustation depends on the exposure time of the surface to sulphide gas environment. Fig. 2 shows the optical images of the tarnished and non-tarnished surface. Cross-sectional image of the tarnish as shown in fig. 2(c) shows a uniform encrustation of about 8-10 µm developed on the surface.

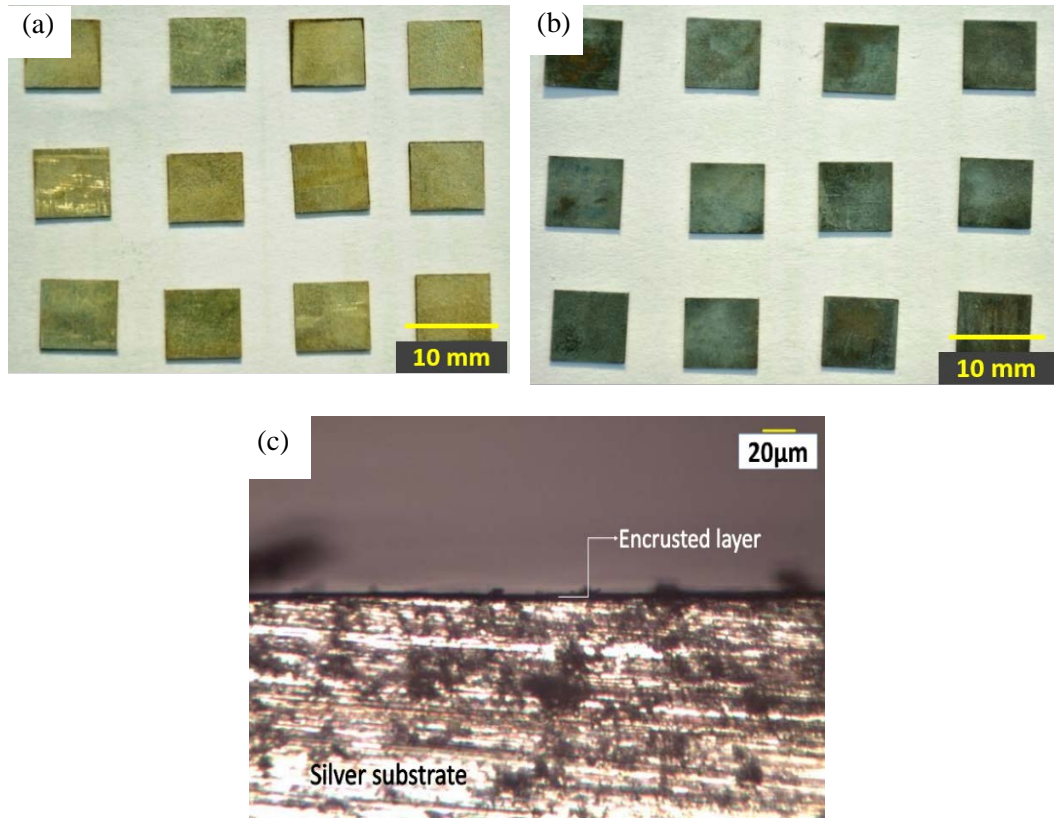

**Fig. 2.** Images of silver samples (a) without tarnish (b) with tarnish (c) Cross section of tarnished surface showing encrustation

*2.2 Experimentation details*

The experimentation was performed using excimer laser (Lambda-Physik) working with KrF chemistry having wavelength 248 nm, pulse width 25 ns, repetition rate 4 Hz, pulse energy (maximum) 650 mJ and rectangular beam profile. Since the output beam was not homogeneous [30], a beam delivery system has been incorporated in the laser as shown in fig. 3 so that a uniform intensity homogenous beam profile can be generated. Fig 4(a) shows the beam intensity profile for both x and y-axis while fig. 4(b) shows schematic of the cleaning process performed on the surface. The laser was operated in constant energy mode throughout the entire experiment.

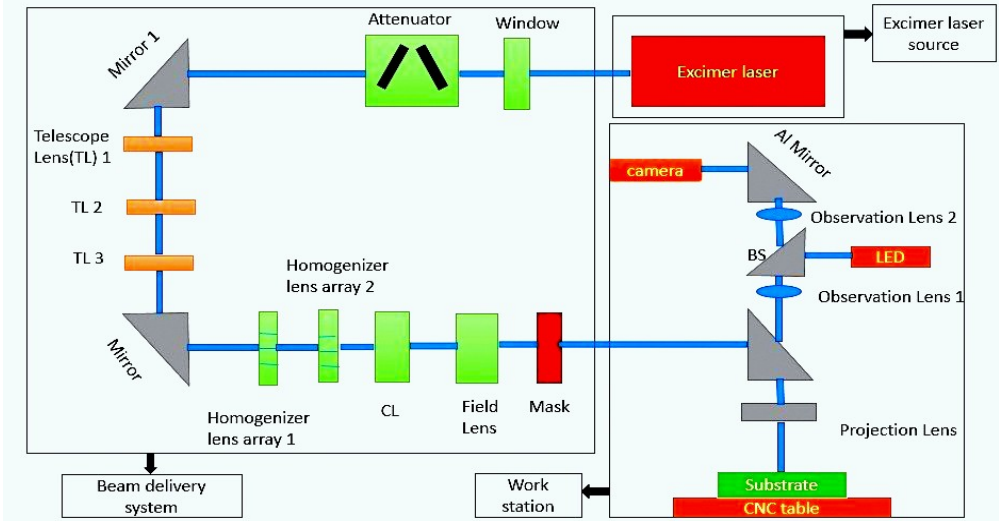

**Fig. 3.** Experimental set up excimer laser used for laser ablation process

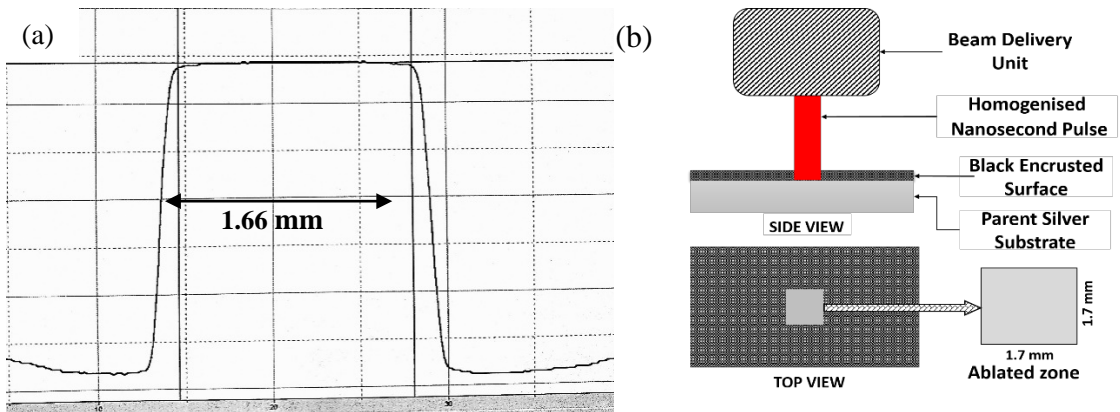

**Fig. 4.** (a) Laser beam profile (b) Schematic of laser cleaning process

In order to perform laser cleaning process, fluence (range: 250-400 mJ/cm$^2$), repetition rate (4 Hz) and numbers of pulses (range: 60-300) were selected based on trial experiments and previous works. These parameters were varied in order to determine their effect on the removal of silver sulphide from the surface. The reason for selecting the lower frequency (4 Hz) can be attributed to the fact that the effect of heat of one pulse on the other can be minimized as well as ablation depth of silver can be controlled [14]. The parametric combinations of the experiment are shown in table 2. After laser ablation, both top surface and cross-sectional surface of the sample were analyzed and characterized.

Table 2 Working parameters for laser cleaning process

| Constant parameters | | Variable parameters | |
|---|---|---|---|
| Ablated area (mm$^2$) | 1.7×1.7 | Laser fluence (mJ/cm$^2$) | 260.48, 286.17, 321.69, 384.85 |
| Repetition rate (Hz) | 4 | | |
| Encrustation thickness | 8-10 µm | No. of pulses | 60, 140, 220, 300 |
| Sample thickness | 2.2 mm | | |

## 3. Results and discussions

*3.1 Effect of laser fluence and number of pulses on sulphur removal*

The effect of laser fluence and number of laser pulses were investigated for quantifying sulphur content on the surface. The initial wt. % of sulphur present on the tarnished surface was around 5.5 wt. % (as shown in table 4). It was observed that the tarnished surface showed an increase in wt. % of sulphur along with carbon, copper and oxygen. After laser cleaning, EDS analysis was done to analyse the wt. % of sulphur. Table 3 present the wt. % of sulphur in the ablated zone with different parametric combinations. The effect of laser fluence and number of pulses on the amount of sulphur present was plotted and presented.

Table 3 Effect of Fluence and number of pulses on wt.% of sulphur

| Frequency Hz | Fluence (mJ/cm$^2$) | Presence of sulphur (wt. %) | | | |
|---|---|---|---|---|---|
| | | 60 pulses | 140 pulses | 220 pulses | 300 pulses |
| 4 | 260.48 | 2.61 | 1.85 | 0.74 | 0.56 |
| | 286.17 | 2.51 | 1.28 | 0.53 | 0.49 |
| | 321.69 | 2.37 | 1.45 | 0.54 | 0.45 |
| | 384.85 | 1.27 | 0.66 | 0.49 | 0.38 |

The effect of fluence on the sulphur content has been shown in fig. 5(a). Laser fluence has a direct role to play in the wt. % of sulphur. From the values of table 3, it is found that the sulphur present on the surface at 260.48 mJ/cm$^2$ with 60 pulses was 2.61 wt. % while sulphur present at 384.85 mJ/cm$^2$ with 300 pulses was 0.38 wt. %. As we increase the fluence keeping number of pulses constant the mean sulphur content decreases. The decrease in sulphur at a fluence of 260.48 mJ/cm$^2$ suggests that the fluence is above the threshold value to cause the dissociation of the bond of silver sulfides. Again it can be said that at 60 pulses the decrease in sulphur is not very drastic, but is uniform. As we increase the number of pulses to 140 and above, we observed the non-uniform changes in the sulphur content. As we know that the mechanism of ablation changes with a change in fluence and intensity [29], so it can be said that the change in fluence value is causing the mechanism of the ablation to change causing the different rate of ablation. Both photochemical, as well as photo-thermal effects, are taking place over here as suggested from the SEM images obtained showing some melting of the substrate surfaces in fig.8(e) and fig.9(b).The optical and thermal properties of the surface are changing after each pulse and with different fluence values. These changes in the properties like absorptivity, reflectivity, conductivity etc. are causing a different amount of heat flow along the body. The surface roughness of the silver sulfide and the silver is also causing the ablation rate to change. Sulfide in the pin holes cannot be removed easily. These sulfides will cause the ablation rate to change. The evaporation threshold of the silver sulfide and the melting threshold of the silver plays a crucial role in the ablation rate [14]. As clear from previous literature [7], the sulphur gets dissociated from silver at around 838 K while the melting point of silver is around 1234 K. A good ablation rate can be achieved if applied fluence can generate temperature within this range. The trend of decrease of sulphur is present in all the cases. At 220 and 300 pulses the sulphur content changes at a slow rate with change in fluence value. Finally, the amount of sulphur in laser treated zone converges to around 0.35-0.55 wt. %.

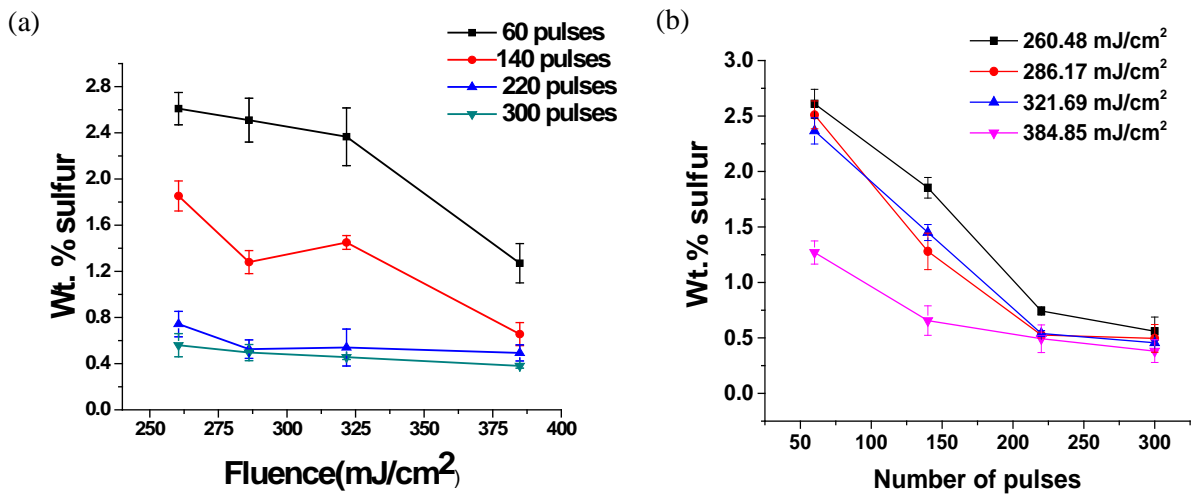

**Fig. 5.** Effect of (a) fluence on wt. % of sulphur (after laser ablation) (b) number of pulses on wt. % of sulphur (after laser ablation).

The effect of number of pulses on the sulphur content has been shown in fig. 5(b). It was observed that 60 number of pulses the ablation rate is more controlled. At 140 number of pulses the ablation rate increases at a faster rate with increase in fluence value. But at 220 pulses again it is getting approximately same for all the fluence values and the fluence increase is not causing any significant variation in the wt. % of sulphur. Overall it can be said that increase in pulse number is causing an increase in ablation and decrease in sulphur content on the surface. At 300 pulses, it can be seen that for the fluence value of 260.48 mJ/cm$^2$ the wt. % of sulphur is at a value of 0.56 and then by increasing the fluence very small change in wt. % of sulphur is taking place. Relatively less amount of sulphur present on it is one of the reason for such a low ablation at 300 pulses. But at a fluence of 384.85 mJ/cm$^2$, the change in sulphur content occurs at a faster rate.The removal of sulphur from the surface at higher fluence is relatively high.So the silver exposed after the sulphur is absorbing the high fluence value.Thus the ablation of silver is also taking place at a higher fluence of 384.85 mJ/cm$^2$.SEM images and EDS analysis (fig. 8(b) and fig. 9(c)) also confirms these findings.

*3.2 Micrographs of the laser cleaned samples*

Optical imaging of the cleaned samples was done using a metallurgical microscope (Zeiss Zoom: Germany). Fig.6 (a) shows some of the lasers cleaned samples with 260.48 mJ/cm$^2$ for a different number of pulses such as 60, 140, 220 and 300 pulses. The boundary of the encrusted samples and cleaned samples is clearly visible. The size of the cleaned surface is around 1.7 mm × 1.7 mm. There are some black patches present within the cleaned zone. These black patches are due to remaining silver sulphide present on the surface. Fig. 6(b) shows the optical images of the laser cleaned zone with 384.85 mJ/cm$^2$ and 220 pulses. The formation of a peripheral rim has taken place for higher fluence value (fig. 6(b)). However, such a rim structure was not observed in case of lower fluence like 260.48 mJ/cm$^2$.The black patches are less prominent in this case. We can say that for higher fluence, stronger shock waves are getting generated at higher fluence and they are causing an increase in ablation rate [14].

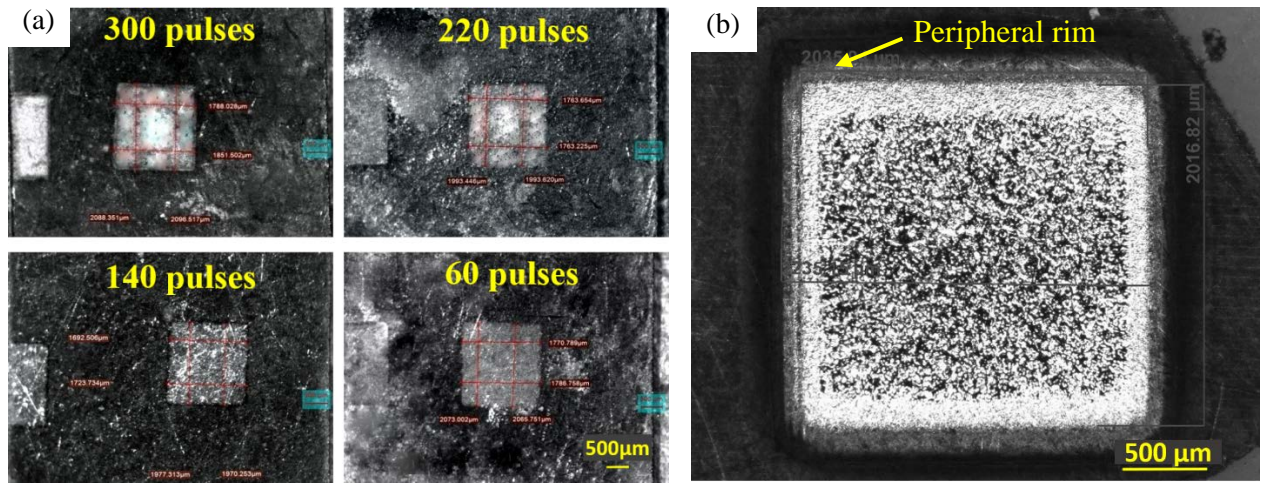

**Fig. 6.** Optical images of some of the laser cleaned at (a) 260.48 mJ/cm$^2$ for a different number of pulses, (b) Optical image of laser cleaned surface at 384.85 mJ/cm$^2$ with 220 pulses

*3.3 Secondary Electron images of the encrusted samples and the laser cleaned samples*

SEM was carried out in order to study the encrusted surface and the cleaned surface. Fig.7 shows the encrusted surface and cleaned surface for 260.48 mJ/cm$^2$ and 300 pulses. It can be seen from fig. 7(a) that the silver was having some initial surface roughness. After sulphur encrustation, we can see that surface bears particulates of silver sulphide and other particulates all over the surface (fig. 7(b)). After laser cleaning, it is generating a uniformly cleaned surface. A uniform cleaned surface can be observed which is assumed to be created by high pressure super quick ablation removal of material. If we look into the laser cleaned surface at higher magnification (fig. 7(e)), we will find that some small granules-like structure present on it. These structures are around 0.5-3 µm in size. These granular structures are unreacted silver sulfide particles present in the pit holes of the silver surface. It can be suggested that some photo-thermal effect is also taking place along with photo-chemical effect as discussed in the previous section. This effect occurs when there is a non- linear laser material interaction occurring in the laser treated zone. These nonlinear interactions can be due to surface roughness or due to change in thermal properties at a higher temperature. Some changes in molecular level may also be the cause of such phenomena. As we know that metallic forces exist between Ag-Ag molecules. These metallic forces vary in their nature from one material to other. In case of silver, the metallic forces are stronger with respect to the ionic compound which is held together with the help of electrostatic forces as in

Ag$_2$S.So some photoionization and dissociation might be the reason for melting of the surface and more sulphur removal at high fluence value.

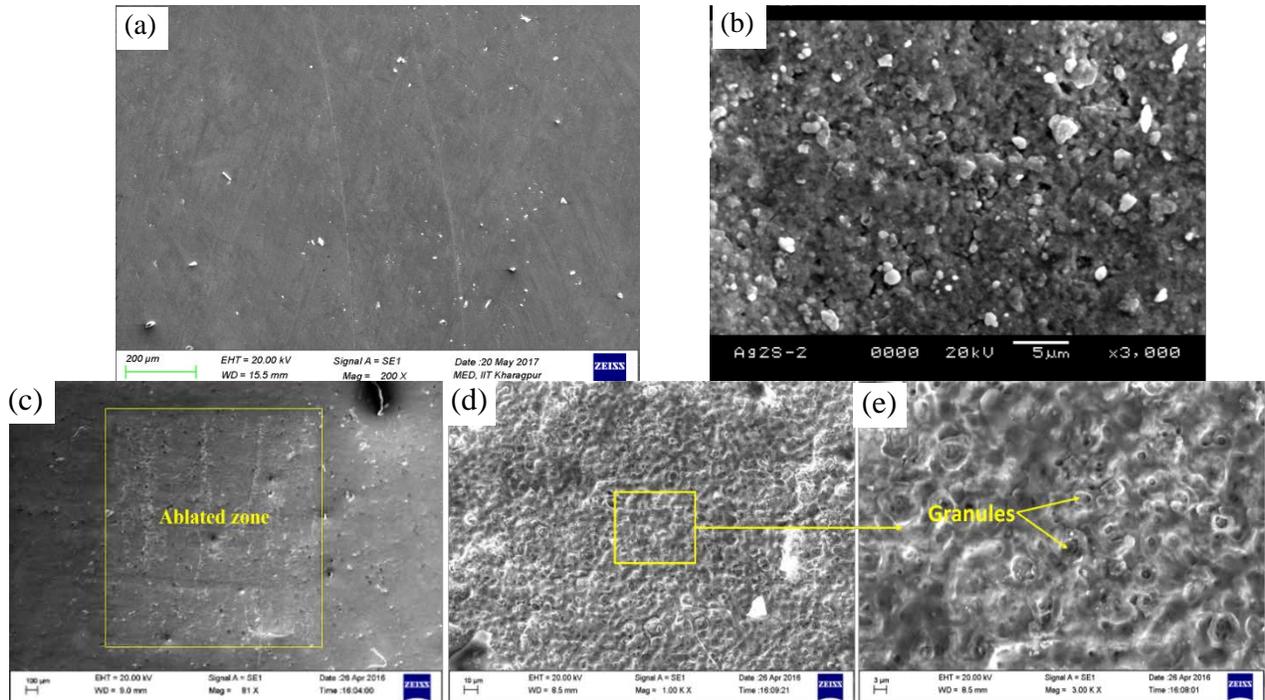

**Fig. 7.** SEM images of (a) silver surface (b) encrusted silver surface (c) laser cleaned surface with 260.48 mJ/cm$^2$ and 300 pulses at 81 X magnification (d) 1KX magnification (e) 3KX magnification.

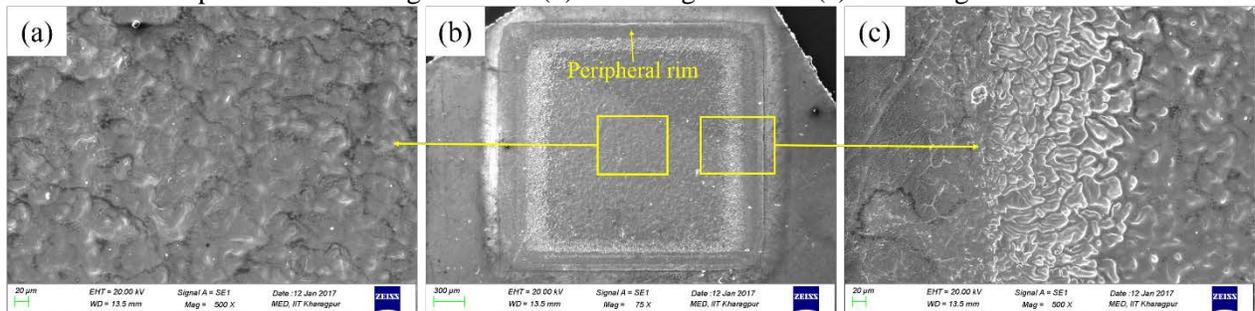

**Fig. 8.** SEM images of the laser cleaned surface at 384.85 mJ/cm$^2$ and 220 pulses showing (a) centre of the laser cleaned zone and (b) overall laser cleaned zone with peripheral rim (c) boundary of the laser cleaned zone.

Fig.8 shows the laser cleaned surface at a higher fluence value of 384.85 mJ/cm$^2$ and 220 pulses. Some peripheral rim type structure is present around the ablated zone as shown in fig. 8(b).This phenomenon is common in laser ablation process due to the free expansion of the superheated pressurized material expulsion that generates a strong shockwave [31,32]The EDS analysis also showed that there is an increased concentration of sulphur and a lower concentration of silver in the peripheral rim structure zone (fig. 9(c, d)).The centre ablated zone as in fig. 8(a) shows that the granular structure has got reduced with respect to the centre zone at a fluence 260.48 mJ/cm$^2$ and 300 pulses as in fig.7(c).A relatively uniform surface has been generated in this case. Fig 8(c) shows the boundary of the ablated zone having a distinct silver surface (offset white) visible with low sulphur.

*3.4 Energy Dispersive X-Ray Spectroscopy Analysis*

Table 4 shows the Energy Dispersive X-Ray Spectroscopy (EDS) report of the sulphur content (in wt. %) of the parent material, encrusted samples and the laser cleaned samples. It is found that the sulphur content has been reduced remarkably. Further, the line EDS was performed to study the change of sulphur content in the ablated and non-ablated zone. Fig. 9(a) and 9(b) shows the line EDS report of silver and sulphur present in the ablated zone for 260.48 mJ/cm$^2$ and 300 pulses while fig. 9(c) and 9(d)) shows the line EDS and EDS area mapping of the sulphur and silver content respectively for 384.85 mJ/cm$^2$ and 220 pulses. It is observed that the sulphur amount has decreased significantly after laser ablation, though the complete removal has not taken place. So, for complete removal, it is suggested that the fluence rate or the number of pulses should be increased. But as we increase the fluence rate we have seen some peripheral rim structure formations as discussed in section 3.3.This peripheral rim has relatively less silver exposed and more sulphur content on its surface(fig 9(c)).Area mapping of the sulphur and silver content (fig. 9(d)) also confirms this finding.

**Table 4** EDS analysis of the parent and laser ablated surface at 260.48 mJ/cm$^2$ and 220 pulses

| Elements | C | O | S | Cu | Ag |
|---|---|---|---|---|---|
| **Uncoated samples** | 1.91 | 0.12 | 0.07 | 0.72 | 97.18 |
| **Coated samples** | 4.04 | 3.86 | 5.56 | 1.68 | 84.87 |
| **After laser ablation** | 1.34 | 1.39 | 0.74 | 0.34 | 96.19 |

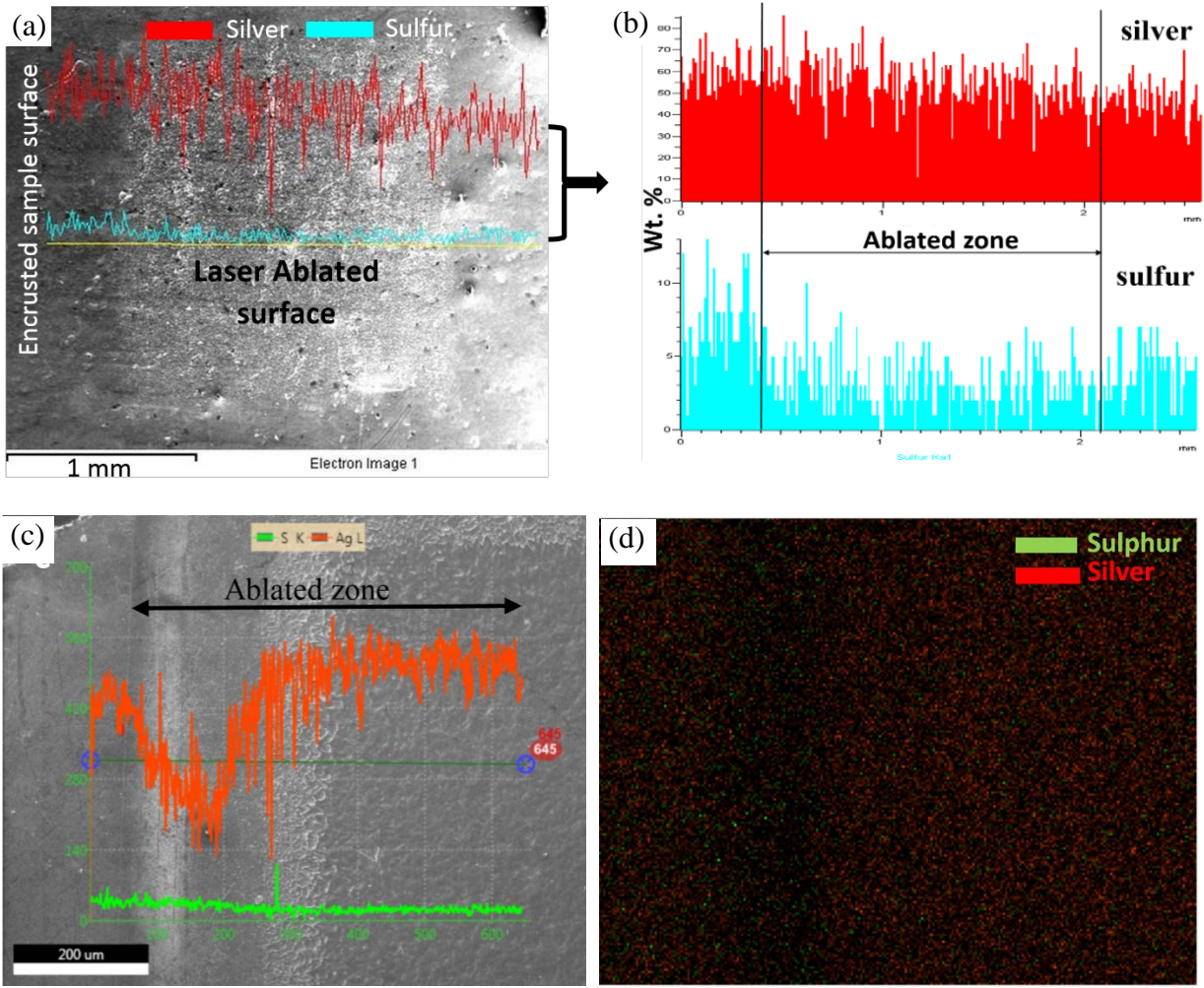

**Fig. 9.** EDS of laser cleaned (a), (b) changes in silver and sulphur amount in encrusted to cleaned surface at 260.48 mJ/cm$^2$ and 300 pulses. (c) Line EDS and (d) Area mapping of laser cleaned surface at 384.85 mJ/cm$^2$ and 220 pulses.

*3.5 Surface topographical analysis*

Three-dimensional contact type profilometer (Taylor-Hobson model-Form Talysurf 50 series, UK) was used to analyse the topography of the laser ablated surface. The topography as in fig. 10 shows the sample cleaned with a fluence of 260.48 mJ/cm$^2$ and 300 pulses (fig. 10(a) and 10(c)) and 384.85 mJ/cm$^2$ and 220 pulses (fig. 10(b) and 10(d)). It is found that 5-8 µm thicker silver sulphide black layer has been removed during the cleaning operation. The surface topography of the ablated zone shows that the ablated zone has a higher roughness than that of the encrusted zone. This happened due to the partial melting and vaporization of the sulphur content from the surface. Some of the process parameters generated a peripheral rim structure beside the ablated zone. The formation of peripheral rim is easily observed in case of fig. 10(b) and 10(d). With the increase in fluence, it was observed that the tendency of formation of peripheral rim got increased. There was a distinct fluence value (321.69 mJ/cm$^2$) after which the peripheral rim structure formation was prominent (fig. 10(b)). Although the beam used was homogenised, yet the ablated area shows more material removal at the centre than the peripheral region. The reason can be attributed to the combined effect of photo-chemical as well as photo-thermal effect taking place in the ablated zone as discussed in section 3.2.

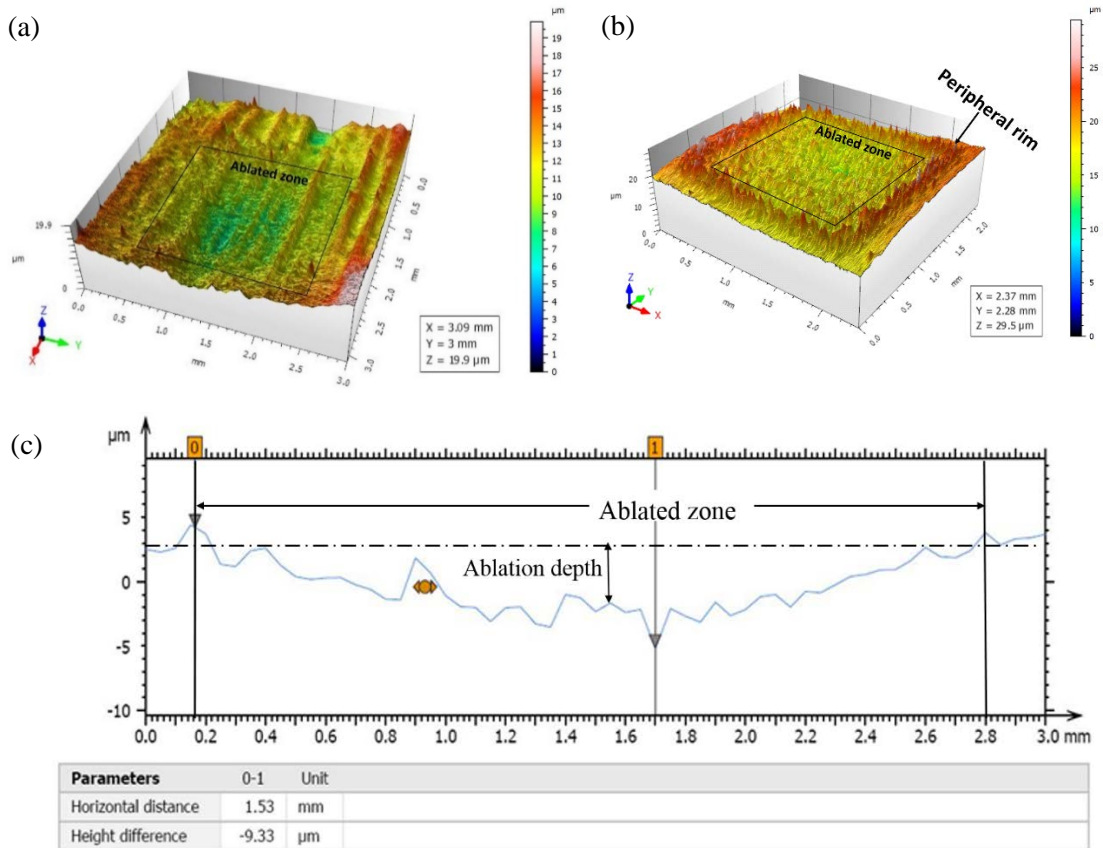

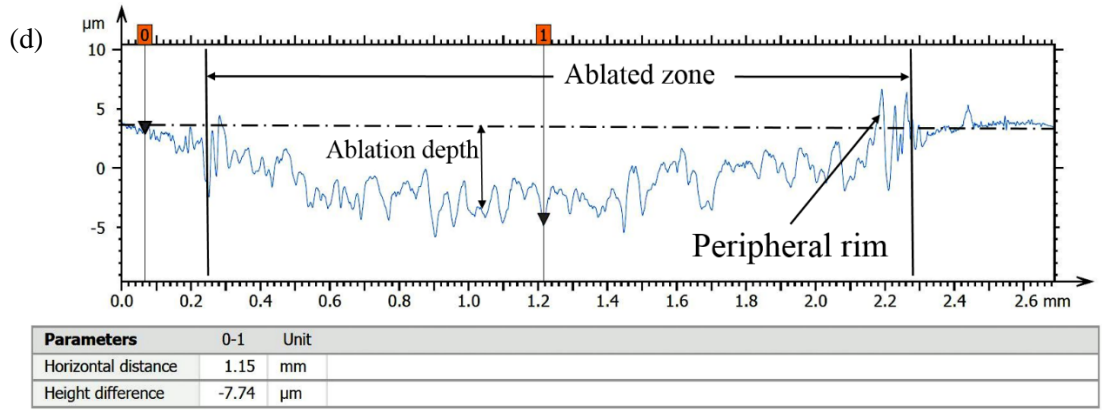

**Fig. 10.** (a) Surface topography of a laser cleaned at (a) 260.48 mJ/cm² and 300 pulses, (b) at 384.85 mJ/cm² and 220 pulses, (c) Cross section profile of the ablated zone for 260.48 mJ/cm² and 300 pulses, (d) Cross section profile of the ablated zone for roughness for 384.85 mJ/cm2 and 220 pulses.

## 3.6 Microhardness Analysis

Vickers microhardness values were measured using Omnitech Semi-Automatic Microhardness Tester. A 50 gm load was applied for a dwell time of 10 seconds. The microhardness values of the laser cleaned areas were taken to compare the change in hardness between laser treated and untreated surface. The hardness was measured at 5 different points of the same location and the average value of the hardness was taken. Fig. 11 shows the change in microhardness from the centre of the ablated zone towards the encrusted sample surface. The uncoated silver showed a microhardness value of 75.12 $HV_{0.05.}$ From the result of microhardness, we can observe that the microhardness difference from cleaned silver to encrusted sample is of about 10 $HV_{0.05.}$ The cleaned surface showed an average hardness value of 96.07 $HV_{0.05}$ while encrusted surface showed an average value of 85.39 $HV_{0.05}$. The change in hardness can be attributed to the partial melting and heating of the cleaned zone causing elemental compositional changes in the treated zone. The microhardness in the centre of the cleaned region is higher than the peripheral part. It can also be verified using the fig.10(c) where it is quite clear that the centre has more ablation depth than the surrounding areas. So more hardness is present at the centre as silver has more hardness than silver sulphide. The periphery of the ablated zone shows microhardness value of around 75 $HV_{0.05}$.

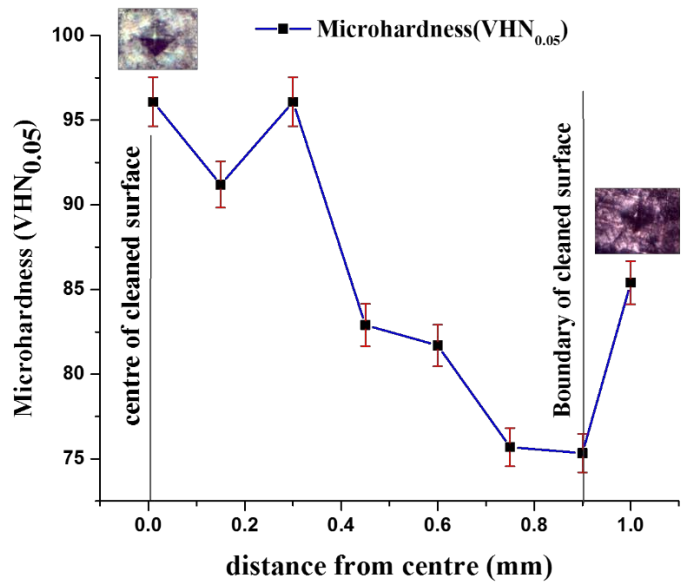

**Fig 11.** The variation of Vicker's microhardness with the position of the ablated surface for 260.48 mJ/cm2 and 300 pulses.

## 4. Conclusion

Following conclusions can be drawn from the cleaning of the sulphide from the silver surface

- Characterisation of the laser cleaned surface showed a significant cleaning of silver sulfide from ~ 6 wt. % to ~0.38 wt. %.
- The fluence value of 260.48 mJ/cm$^2$ is well above the threshold value for removal of sulphide from the encrusted surface.
- A maximum removal was observed for 384.85 mJ/cm$^2$ and 220 pulses where the remaining sulphur present on the surface was around 0.38 wt. %.
- The SEM image analysis showed that peripheral rim was observed after a fluence value of 286.17 mJ/cm$^2$. This peripheral rim became more prominent with an increase in the value of fluence as well as number of pulses. The higher fluence of 321.69 and 384.85

mJ/cm$^2$ showed more sulphur removal with low pulses but peripheral rim formations pose a drawback of such high fluence values.

- Some amount of melting was also observed for higher fluence values of 321.69 and 384.85 mJ/cm$^2$. So an online monitoring system with colour change sensors and the feedback loop is crucial in case of preventing damage to the base material.

- At 300 pulses all the fluence values showed approximately same wt. % of sulphur present on the cleaned surface. This suggests that the higher number of pulses can be used to clean the surface even with a low fluence but the value of the fluence must be above the threshold fluence required for the laser cleaning operation. A higher number of pulses will require more cleaning time, so a tradeoff is required between the precision required and the cleaning time.

- The topography of the ablated surface showed a removal of 5-10 µm of the sulphide layer from the surface. But the roughness of the ablated surface has got increased.

- Microhardness value showed a change of microhardness of the silver surface from 75.12 HV$_{0.5}$ (uncoated parent silver) to 96.07 HV$_{0.5}$ (centre of ablated zone).

- Finally, it can be said that excimer laser can definitely be used to clean the silver components in an electronic circuit without affecting the other components as it provides a clean source of energy with sufficient intensity.

**References**

[1]   C. Theodorakopoulos, V. Zafiropulos, Laser cleaning applications for religious objects, European Journal of Science and Theology. 1 (2005) 63–76.
[2]   J.S. Pozo-Antonio, T. Rivas, M.P. Fiorucci, A.J. López, A. Ramil, Effectiveness and harmfulness evaluation of graffiti cleaning by mechanical, chemical and laser procedures on granite, Microchem. J. 125 (2016) 1–9. doi:10.1016/j.microc.2015.10.040.
[3]   A.C. Tam, H.K. Park, C.P. Grigoropoulos, Laser cleaning of surface contaminants, Appl.

Surf. Sci. 127–129 (1998) 721–725. doi:10.1016/S0169-4332(97)00788-5.

[4] P.V. Maravelaki, V. Zafiropulos, V. Kilikoglou, M. Kalaitzaki, C. Fotakis, Laser-induced breakdown spectroscopy as a diagnostic technique for the laser cleaning of marble, Spectrochim. Acta Part B At. Spectrosc. 52 (1997) 41–53. doi:10.1016/S0584-8547(96)01573-X.

[5] M. Sokhan, P. Gaspar, D.S. McPhail, A. Cummings, L. Cornish, D. Pullen, F. Hartog, C. Hubbard, V. Oakley, J.F. Merkel, Initial results on laser cleaning at the Victoria & Albert Museum, Natural History Museum and Tate Gallery, J. Cult. Herit. 4 (2003) 230–236. doi:10.1016/S1296-2074(02)01219-0.

[6] A. Sansonetti, M. Colella, P. Letardi, B. Salvadori, J. Striova, Laser cleaning of a nineteenth-century bronze sculpture: *In situ* multi-analytical evaluation, Stud. Conserv. 60 (2015) S28–S33. doi:10.1179/0039363015Z.000000000204.

[7] I. Turovets, M. Maggen, a Lewis, Cleaning of daguerreotypes with an excimer laser, Stud. Conserv. 43 (1998) 89–100. doi:10.1179/sic.1998.43.2.89.

[8] G.S. Senesi, I. Carrara, G. Nicolodelli, D.M.B.P. Milori, O. De Pascale, Laser cleaning and laser-induced breakdown spectroscopy applied in removing and characterizing black crusts from limestones of Castello Svevo, Bari, Italy: A case study, Microchem. J. 124 (2016) 296–305. doi:10.1016/j.microc.2015.09.011.

[9] W. Kautek, S. Pentzien, P. Rudolph, J. Krüger, E. König, Laser interaction with coated collagen and cellulose fibre composites: Fundamentals of laser cleaning of ancient parchment manuscripts and paper, Appl. Surf. Sci. 127–129 (1998) 746–754. doi:10.1016/S0169-4332(97)00735-6.

[10] P. Maravelaki-Kalaitzaki, V. Zafiropulos, C. Fotakis, Excimer laser cleaning of encrustation on Pentelic marble: Procedure and evaluation of the effects, Appl. Surf. Sci. 148 (1999) 92–104. doi:10.1016/S0169-4332(99)00125-7.

[11] S. Klein, F. Fekrsanati, J. Hildenhagen, K. Dickmann, H. Uphoff, Y. Marakis, V. Zafiropulos, Discoloration of marble during laser cleaning by Nd:YAG laser wavelengths, Appl. Surf. Sci. 171 (2001) 242–251. doi:10.1016/S0169-4332(00)00706-6.

[12] J. Kolar, M. Strlič, D. Müller-Hess, A. Gruber, K. Troschke, S. Pentzien, W. Kautek, Laser cleaning of paper using Nd:YAG laser running at 532 nm, J. Cult. Herit. 4 (2003) 185–187. doi:10.1016/S1296-2074(02)01196-2.

[13] D.A. Wesner, M. Mertin, F. Lupp, E.W. Kreutz, Cleaning of copper traces on circuit boards with excimer laser radiation, Appl. Surf. Sci. 96–98 (1996) 479–483. doi:10.1016/0169-4332(95)00499-8.

[14] J. Zhang, Y. Wang, P. Cheng, Y.L. Yao, Effect of pulsing parameters on laser ablative cleaning of copper oxides, J. Appl. Phys. 99 (2006). doi:10.1063/1.2175467.

[15] C. Seo, D. Ahn, D. Kim, Removal of oxides from copper surface using femtosecond and nanosecond pulsed lasers, Appl. Surf. Sci. 349 (2015) 361–367. doi:10.1016/j.apsusc.2015.05.011.

[16] A. Kearns, C. Fischer, K.G. Watkins, M. Glasmacher, H. Kheyrandish, A. Brown, W.M. Steen, P. Beahan, Laser removal of oxides from a copper substrate using Q-switched Nd:YAG radiation at 1064 nm, 532 nm and 266 nm, Appl. Surf. Sci. 127–129 (1998) 773–780. doi:10.1016/S0169-4332(97)00741-1.

[17] A.C. Tam, W.P. Leung, W. Zapka, W. Ziemlich, Laser-cleaning techniques for removal of surface particulates, J. Appl. Phys. 71 (1992) 3515–3523. doi:10.1063/1.350906.

[18] K.G. Watkins, C. Curran, J.-M. Lee, Two new mechanisms for laser cleaning using

Nd:YAG sources, J. Cult. Herit. 4 (2003) 59–64. doi:10.1016/S1296-2074(02)01229-3.

[19] J.-M. Lee, J.-E. Yu, Y.-S. Koh, Experimental study on the effect of wavelength in the laser cleaning of silver threads, J. Cult. Herit. 4 (2003) 157–161. doi:10.1016/S1296-2074(02)01192-5.

[20] A. Singh, A. Choubey, M.H. Modi, B.N. Upadhyaya, S.M. Oak, G.S. Lodha, S.K. Deb, Cleaning of carbon layer from the gold films using a pulsed Nd:YAG laser, Appl. Surf. Sci. 283 (2013) 612–616. doi:10.1016/j.apsusc.2013.06.157.

[21] M.J.J. Schmidt, L. Li, J.T. Spencer, An investigation into the feasibility and characteristics of using a 2.5 kW high power diode laser for paint stripping, J. Mater. Process. Technol. 138 (2003) 109–115. doi:10.1016/S0924-0136(03)00057-8.

[22] F. Brygo, A. Semerok, R. Oltra, J.M. Weulersse, S. Fomichev, Laser heating and ablation at high repetition rate in thermal confinement regime, Appl. Surf. Sci. 252 (2006) 8314–8318. doi:10.1016/j.apsusc.2005.11.036.

[23] Y.K. Madhukar, S. Mullick, D.K. Shukla, S. Kumar, A.K. Nath, Effect of laser operating mode in paint removal with a fiber laser, Appl. Surf. Sci. 264 (2013) 892–901. doi:10.1016/j.apsusc.2012.10.193.

[24] M.W. Turner, P.L. Crouse, L. Li, A.J.E. Smith, Investigation into CO2 laser cleaning of titanium alloys for gas-turbine component manufacture, Appl. Surf. Sci. 252 (2006) 4798–4802. doi:10.1016/j.apsusc.2005.06.061.

[25] Y.C. Guan, G.K.L. Ng, H.Y. Zheng, M.H. Hong, X. Hong, Z. Zhang, Laser surface cleaning of carbonaceous deposits on diesel engine piston, Appl. Surf. Sci. 270 (2013) 526–530. doi:10.1016/j.apsusc.2013.01.075.

[26] W. Liu, X. Qing, M. Li, L. Liu, H. Zhang, Supercritical CO2 cleaning of carbonaceous deposits on diesel engine valve, Procedia CIRP. 29 (2015) 828–832. doi:10.1016/j.procir.2015.02.014.

[27] L. Yue, Z. Wang, L. Li, Modeling and simulation of laser cleaning of tapered micro-slots with different temporal pulses, Opt. Laser Technol. 45 (2013) 461–468. doi:10.1016/j.optlastec.2012.05.036.

[28] W. Zhang, Y.L. Yao, K. Chen, Modelling and Analysis of UV Laser Micromachining of Copper, Int J Adv Manuf Technol. 18 (2001) 323–331. doi:10.1007/s001700170056.

[29] S. Siano, Principles of Laser Cleaning in Conservation, Handb. Use Lasers Conserv. Conserv. Sci. 7 (2007) 1–26.

[30] M. Technology, Micromachining of silicon with excimer laser in air and water medium Alok Kuma, 21 (2010) 42–53. doi:10.1504/IJMTM.2010.034285.

[31] A.P. Singh, A. Kapoor, K.N. Tripathi, G.R. Kumar, Laser damage studies of silicon surfaces using ultra-short laser pulses, Opt. Laser Technol. 34 (2002) 37–43. doi:10.1016/S0030-3992(01)00090-1.

[32] D. Von Der Linde, K. Sokolowski-Tinten, Physical mechanisms of short-pulse laser ablation, Appl. Surf. Sci. 154 (2000) 1–10. doi:10.1016/S0169-4332(99)00440-7.